\documentclass[english]{article}
\usepackage[T1]{fontenc}
\usepackage[latin9]{inputenc}
\usepackage{CJK}
\usepackage{setspace}
\usepackage{esint}
\usepackage{color}
\usepackage{babel}

\date{}
\makeatletter
\newcommand{\lyxaddress}[1]{
\par {\raggedright #1
\vspace{1.4em}
\noindent\par}
}

\definecolor{BLACK}{gray}{0}
\definecolor{WHITE}{gray}{1}
\definecolor{RED}{rgb}{1,0,0}
\definecolor{GREEN}{rgb}{0,1,0}
\definecolor{BLUE}{rgb}{0,0,1}
\definecolor{CYAN}{cmyk}{1,0,0,0}
\definecolor{MAGENTA}{cmyk}{0,1,0,0}
\definecolor{YELLOW}{cmyk}{0,0,1,0}
\definecolor{lightgray}{gray}{0.7}

\newcommand{\heq}{{\ {\hat =}\ }}

\makeatother

\begin{document}

\title{Generalized Kerr/CFT correspondence with electromagnetic field}

\author{ Cheng-Yong Zhang$\ensuremath{^{1}}$%
\thanks{zhangcy@sjtu.edu.cn%
}, Yu Tian$\ensuremath{^{2,4}}$%
\thanks{ytian@ucas.ac.cn%
}, Xiao-Ning Wu$\ensuremath{^{3,4,5}}$%
\thanks{wuxn@amss.ac.cn%
}}

\maketitle

\begin{singlespace}
\lyxaddress{\begin{center}
1. Department of Physics and Astronomy, Shanghai Jiao Tong University,
Shanghai 200240, China\\
2. School of Physics, University of Chinese Academy of Sciences,
Beijing 100049, China\\
3. Institute of Mathematics, Academy of Mathematics and System Science,
Chinese Academy of Sciences, Beijing 100190, China\\
4. State Key Laboratory of Theoretical Physics, Institute of Theoretical
Physics, Chinese Academy of Sciences, Beijing 100190\\
5. Hua Loo-Keng Key Laboratory of Mathematics, CAS, Beijing 100190,
China
\par\end{center}}
\end{singlespace}
\begin{abstract}
The electrovac axisymmetric extreme isolated horizon/CFT correspondence
is considered.{ By expansion techniques under the Bondi-like coordinates,}
it is {proved} that the near horizon geometry of electrovac
axisymmetric extreme isolated horizon is unique. {Furthermore,
explicit coordinate transformation between the Bondi-like coordinates
and the Poincare-type coordinates for the near horizon metric of the extreme Kerr-Newmann
spacetime is found.}{ Based on these analyses and the thermodynamics of the
isolated horizon, then,} the Kerr/CFT correspondence is generalized to nonstationary
extreme black holes with electromagnetic fields.
\end{abstract}

\section{Introduction}

The discovery of Bekenstein-Hawking entropy was a great break through
in theoretical physics in the last decades, but the microscopic origin
of it still remains unclear. In \cite{Entropy_5D,Entropy_4D}, the
entropy of 5-dimensional and 4-dimensional extreme black holes in
string theory were derived by counting the degeneracy of BPS soliton
bound states. Based on the demonstration that any consistent theory
of quantum gravity on $AdS_{3}$ is holographically dual to a 2-dimensional
conformal field theory (CFT) did not invoke string theory\cite{AdS3_CFT},
entropy of the black holes whose near horizon geometry is locally
$AdS_{3}$ was counted in a dual 2-dimensional conformal field theory
without using string theory or supersymmetry\cite{CFT_1}.

Paralleling to the $AdS_{3}$/$CFT_{2}$ duality, the correspondence
between the near horizon extreme Kerr (NHEK) geometry and CFT was
found in \cite{Kerr_CFT}, via a near horizon limiting procedure.
Given the suitably-chosen boundary conditions at the asymptotic infinity
of NHEK, they found that the asymptotic symmetry group (ASG) {extends to}
a Virasoro algebra {upon Noether charge realizations}. The central charge of this algebra was calculated.
Then, with the Frolov-Thorne temperature and the Cardy formula, the
microscopic entropy of extreme Kerr was computed{, which} is found exactly
the same as the macroscopic Bekenstein-Hawking entropy of the black hole. Thus the extreme Kerr black hole is holographically
dual to a chiral 2-dimensional CFT.

This duality has been generalized to many situations, including other
black holes, diverse dimensions, supergravities etc\cite{AOT_1KK,BMPV,Godel,Gauge_SRG,Heterotic,Central_Charges,divers_dim,LMPV_SRG,Wald_Entropy,Min_SRG,RN_CFT}.
{ For a nice review, see \cite{Compere,BKLS}.}
We would like to emphasize that \cite{KN_CFT} generalized the
Kerr/CFT correspondence to the Kerr-Newman-AdS-dS/CFT correspondence
in four dimensions since it will be closely related to this paper. Nevertheless, {most of}
the situations that have been considered are all stationary.
In fact, \cite{IH_CFT} has generalized the Kerr/CFT correspondence
to the vacuum extreme isolated horizon/CFT correspondence. {It is well-known that all stationary horizon satisfy the defination of isolated horizon {(IH)}\cite{Ashtekar_1, IH_DH}, but IH contains some non-stationary cases\cite{Lewandowski, LP02}.}
{It should also be emphasized that the spacetime itself in the IH framework does not need to be axisymmetric,
though the inner horizon data of it is axisymmetric.}

On the other hand, based on the result of Lewandowski and Pawlowski\cite{Uniq_Eextre_IH},
we know that the {inner geometry} of axisymmetric electrovac extreme
{IH} coincide with the extreme Kerr-Newman event
horizon. Thus it is natural to ask whether this correspondence
could be generalized further to electrovac extreme IH/CFT correspondence.

The paper is organized as follows. In section 2, we review the properties
of Kerr-Newman spacetime. In section 3, the  isolated horizon is introduced
and we get the metric of its near horizon limit {by expansion techniques under the Bondi-like coordinates}.
We find that the near horizon geometry of axisymmetric electrovac extreme IH is the same as that
of the extreme Kerr-Newman spacetime. In section 4, we discuss the electrovac extreme IH/CFT
correspondence. Section 5 is {devoted to} the summary and some discussions.

\section{Review of Kerr-Newman spacetime}

The Kerr-Newman (KN) solution of Einstein equation describes the stationary
and axisymmetric rotational charged black hole. The metric of KN spacetime
in Boyer-Lindquist coordinates is

\begin{eqnarray}
ds^{2} & = & -\frac{\Delta}{\rho^{2}}(d\hat{t}-a\sin^{2}\theta d\hat{\phi})^{2}+\frac{\rho^{2}}{\Delta}d\hat{r}^{2}+\rho^{2}d\theta^{2}\label{eq:Boyer}\\
 &  & +\frac{\sin^{2}\theta}{\rho^{2}}(ad\hat{t}-(\hat{r}^{2}+a^{2})d\hat{\phi})^{2}\nonumber
\end{eqnarray}
with
\begin{eqnarray}
\rho^{2} & = & \hat{r}^{2}+a^{2}\cos^{2}\theta,\nonumber \\
\Delta & = & \hat{r}^{2}-2M\hat{r}+a^{2}+Q^{2},\\
Q^{2} & = & Q_{e}^{2}+Q_{m}^{2}.\nonumber
\end{eqnarray}
Here $Q_{e}$ is the electric charge and $Q_{m}$ the magnetic charge.

The electromagnetic potential and field strength of KN are
\begin{eqnarray}
A & = & -\frac{Q_{e}\hat{r}}{\rho^{2}}\left[d\hat{t}-a\sin^{2}\theta d\hat{\phi}\right]-\frac{Q_{m}\cos\theta\hat{r}}{\rho^{2}}\left[ad\hat{t}-(\hat{r}^{2}+a^{2})d\hat{\phi}\right],\\
F & = & \frac{Q_{e}(\hat{r}^{2}-a^{2}\cos^{2}\theta)+2Q_{m}\hat{r}a\cos\theta}{\rho^{4}}\left[d\hat{t}-a\sin^{2}\theta d\hat{\phi}\right]\wedge d\hat{r}\\
 &  & +\frac{Q_{m}(\hat{r}^{2}-a^{2}\cos^{2}\theta)-2Q_{e}\hat{r}a\cos\theta}{\rho^{4}}\sin\theta d\theta\wedge\left[ad\hat{t}-(\hat{r}^{2}+a^{2})d\hat{\phi}\right].\nonumber
\end{eqnarray}

The outer horizon of KN is at $r_{+}=M+\sqrt{M^{2}-(a^{2}+Q^{2})}$.
The Hawking temperature
\begin{equation}
T_{H}=\frac{r_{+}^{2}-(a^{2}+Q^{2})}{4\pi r_{+}(r_{+}^{2}+a^{2})}.\label{eq:Temperature}
\end{equation}

For extreme KN black hole, $T_{H}=\kappa=0$. We have $M^{2}=a^{2}+Q^{2}.$
The outer horizon then becomes $r_{+}=M$, which coincides with the
inner horizon, and the area
\begin{equation}
A=4\pi(M^{2}+a^{2}).
\end{equation}
The Bekenstein-Hawking entropy at extremity is
\begin{equation}
S_{BH}=\frac{A}{4}=\pi(a^{2}+M^{2}).\label{eq:entrop_KN}
\end{equation}
We consider only the extreme KN black hole from now on.

To find the near horizon limit geometry of extreme KN, the following coordinate transformation can be introduced \cite{Horowitz}:
\begin{eqnarray}
\hat{r} & = & M+\epsilon r_{0}r,\nonumber \\
\hat{t} & = & \frac{r_{0}}{\epsilon}t,\label{eq:limit_transf}\\
\hat{\phi} & = & \phi+\Omega_{H}\frac{r_{0}}{\epsilon}t,\nonumber
\end{eqnarray}
where $r_{0}^{2}=M^{2}+a^{2}$ and $\Omega_{H}\equiv\frac{a}{r_{+}^{2}+a^{2}}$
is the angular velocity of the horizon. After taking the limit $\epsilon\rightarrow0$,
metric (\ref{eq:Boyer}) becomes
\begin{eqnarray}
ds^{2} & = & (M^{2}+a^{2}\cos^{2}\theta)(-r^{2}dt^{2}+\frac{dr^{2}}{r^{2}}+d\theta^{2})\label{eq:NHEKN}\\
 &  & +\frac{(M^{2}+a^{2})^{2}\sin^{2}\theta}{M^{2}+a^{2}\cos^{2}\theta}(d\phi+\frac{2Ma}{M^{2}+a^{2}}rdt)^{2}.\nonumber
\end{eqnarray}
Besides, the near horizon {limit of the} electromagnetic field becomes
\begin{eqnarray}
A & = & f(\theta)(d\phi+\frac{2aM}{M^{2}+a^{2}}rdt),\label{eq:gauge_field}\\
f(\theta) & = & \frac{(M^{2}+a^{2})[Q_{e}(M^{2}-a^{2}\cos^{2}\theta)+2Q_{m}aM\cos\theta]}{2(M^{2}+a^{2}\cos^{2}\theta)Ma}.\nonumber
\end{eqnarray}
\inputencoding{latin9}%
\selectlanguage{english}%

According to \cite{KN_CFT}, the asymptotic symmetries of near horizon geometry can be worked out for some suitable boundary conditions. All asymptotic Killing vectors form an algebra and can be expressed as linear combinations of the following bases,
\begin{eqnarray}
\xi_{n} = \epsilon_{n}(\phi)\partial_{\phi}-r\epsilon'_{n}(\phi)\partial_{r}
\end{eqnarray}
with mode $\epsilon_{n}(\phi)=-e^{-in\phi}$. Considering the charge realization of the above algebra\cite{KN_CFT}, one can get a Virasoro algebra with a central charge
\begin{eqnarray}
c = 12Ma=12J.
\end{eqnarray}
Besides, the asymptotic symmetries of KN spacetime also include a $U(1)$ gauge
transformation. However, it has no contribution for the central charge.

Following \cite{Kerr_CFT}, the Frolov-Thorne vacuum for the extreme
KN spacetime was adopted. The quantum fields can be expanded in eigenmodes {with} energy $\omega$ and angular momentum $m$, i.e.
$e^{-i\omega\hat{t}+im\hat{\phi}}$. Here $\hat{t},\hat{\phi}$ are the
Boyer-Lindquist coordinates. In Poincare-type coordinates (\ref{eq:NHEKN}),
we get
\begin{equation}
e^{-i\omega\hat{t}+im\hat{\phi}}=e^{-in_{R}t+in_{L}\phi}.\label{eq:mode}
\end{equation}
 Comparing the coefficients in (\ref{eq:mode}), we get the left and
right charges associated to $\partial_{\phi}$ and $\partial_{t}$
in the near horizon region\cite{divers_dim}.
\begin{eqnarray}
n_{L}=m & , & n_{R}=\frac{(r_{h}^{2}+a^{2})\omega-am}{\epsilon r_{h}}.
\end{eqnarray}
 The Boltzmann factor observed by a canonical observer can be reexpressed
as
\begin{equation}
e^{-\frac{\omega-m\Omega_{H}}{T_{H}}}=e^{-\frac{n_{R}}{T_{R}}-\frac{n_{L}}{T_{L}}}.
\end{equation}
 So the dimensionless left and right temperatures are
\begin{eqnarray}
T_{L}=\frac{T_{H}}{a(M^{2}+a^{2})^{-1}-\Omega_{H}} & , & T_{R}=\frac{(M^{2}+a^{2})T_{H}}{\epsilon M}
\end{eqnarray}
with $T_{H}$ given by (\ref{eq:Temperature}). In the extreme limit,
we get
\begin{eqnarray}
T_{L}=\frac{M^{2}+a^{2}}{4\pi Ma} & , & T_{R}=0.
\end{eqnarray}
This means that the quantum fields outside the horizon are thermally
distributed with temperature $T_{L}$, not in a pure state\cite{Kerr_CFT}.

According to the Cardy formula, the microscopic entropy of the dual
CFT to extreme IH is
\begin{equation}
S_{micro}=\frac{\pi^{2}}{3}cT_{L}=\pi(M^{2}+a^{2}).
\end{equation}
 This is just the same as the macroscopic Bekenstein-Hawking entropy
(\ref{eq:entrop_KN}) of extremal KN black hole.

Now we have reviewed the properties of KN. In the following sections,
we will prove that the near horizon geometry of electrovac extreme
isolated horizons is the same as that of the extreme KN. Then the calculation
of microscopic origin of the Bekenstein-Hawking entropy of electrovac
extreme IHs can be reduced to that of extreme KN.

\section{The near horizon geometry of extreme isolated horizons}

{In last section, we briefly review the Kerr/CFT correspondence. It is clear that the near horizon limit of space-time geometry places an important rule in such correspondence. For stationary black holes, the classification of near horizon limit has been reviewed in \cite{KL13}. In this section,} in order to generalize the Kerr/CFT correspondence to more general
cases, we consider {the near horizon limit of space-time which contains} Ashtekar's isolated horizon\cite{IH_DH}. It has been
proved that all the stationary horizon belongs to IH, including the
Schwarzchild and Kerr horizon. It also contains many nonstationary
cases\cite{Lewandowski, LP02}.

The definition of isolated horizon can be find in \cite{IH_DH}. Roughly speaking, IH is a null hypersurface $\Delta$ with a pair $(h,D)$,
where $h$ is the induced metric tensor and $D$ the induced derivative
operator.
By definition, the generator $l$ of $\Delta$ is shear-free because
of the Raychaudhuri equation and $h_{ab}$ and $D$ are preserved
by $l$. There also exists a rotation one form potential $\omega_{a}$
satisfying $D_{a}l^{b}\heq\omega_{a}l^{b}$ (Here ``$\heq$''
means equality holds only on the horizon $\Delta$). By definition, $\omega_a$ satisfies
\begin{equation}
\mathcal{L}_{l}\omega_{a}\heq 0,\label{eq:1form_poten}
\end{equation}
i.e. $\omega_{a}$ is time independent.

Now we choose the Bondi-like coordinates to describe the near horizon
geometry of IH \cite{IH_DH,Bondi-coor,GeoIH}. In this coordinate system,
we have a complex null tetrad $\{n,l,m,\bar{m}\}$ which could be
expanded as
\begin{eqnarray}
n & = & \partial_{r},\nonumber \\
l & = & \partial_{t}+U\partial_{r}+X\partial_{\zeta}+\bar{X}\partial_{\bar{\zeta}},\nonumber \\
m & = & W\partial_{r}+\xi\partial_{\vartheta}+\zeta\partial_{\bar{\vartheta}},\label{eq:tetrad}\\
\bar{m} & = & \bar{W}\partial_{r}+\bar{\xi}\partial_{\bar{\vartheta}}+\bar{\zeta}\partial_{\vartheta}.\nonumber
\end{eqnarray}
The vectors $l,m,\bar{m}$ span the tangent space to $\Delta$ and
$l$ is the generator of $\Delta$. The metric could be written as
\begin{equation}
g^{ab}=-l^{a}n^{b}-n^{a}l^{b}+m^{a}\bar{m}^{b}+m^{b}\bar{m}^{a}.\label{m1}
\end{equation}

To get the near horizon {limit} of IH, the near horizon behavior
of the metric in (\ref{m1}) should be worked out. We will
complete this work by using the Newman-Penrose (NP) formalism\cite{SKMHH}.
Here the Bondi gauge $\nabla_{n}(n,l,m,\bar{m})=0$ is chosen, which
means that the tetrad is parallel transported along $n$. The connection
between the unknown functions in (\ref{eq:tetrad}) and the spin coefficients
in NP formalism will be built first.

Because $(l,\ m,\ {\bar m})|_{\Delta}$ are tangent vectors on $\Delta$, from eq.(\ref{eq:tetrad}), it means
\begin{equation}
U\heq X\heq W\heq 0\label{eq:UXW0}.
\end{equation}
Under some suitable rotation, $\{m^{a},\bar{m}^{a}\}$ could always
be chosen such that $\mathcal{L}_{l}m^{a}\heq 0$. The chosen gauge
then imply that the spin coefficients {satisfy}\footnote{The notations adopted here are in accord with \cite{SKMHH}. For example,
$-\tau=l_{a;b}m^{a}n^{b}$, $\pi=n_{a;b}\bar{m}^{a}l^{b}$, $\Psi_{0}=C_{abcd}l^{a}m^{b}l^{c}m^{d}$,
$\Phi_{10}=\frac{1}{2}R_{42}$, etc..}
\begin{eqnarray}
\tau=\nu={\gamma=\alpha+\bar{\beta}-\pi=\mu-\bar{\mu}=0}.\label{eq:spin_coeff}
\end{eqnarray}
The Raychaudhuri equations and the definition of IH lead to\cite{GeoIH}
\begin{eqnarray}
\Psi_{0}\ \hat{=}\ 0,\quad\Psi_{1}\ \hat{=}\ 0,\quad\rho\ \hat{=}\ 0,\quad\sigma\ \hat{=}\ 0.\label{P01}
\end{eqnarray}

In the existence of electromagetic field, Raychaudhuri equations also
give
\begin{equation}
F{}_{lm}\heq 0.\label{eq:EM_condition}
\end{equation}
In terms of NP notations, the above result can be expressed as $\Phi_{0}\heq 0$.
This leads to $\Phi_{10}\heq 0$ because of the Einstein equation and the definition of energy-momentum tensor of the Maxwell field.

Based on the results of \cite{GeoIH}, the extreme condition of IH implies
\begin{equation}
\varepsilon\ \hat{=}\ 0.
\end{equation}

After these analyses, the near horizon metric and electromagnetic
field can be solved. The first Cartan structure equations lead to\cite{IH_CFT}
\begin{eqnarray}
\partial_{r}W & = & \bar{\pi}-\mu W-\bar{\lambda}\bar{W}\ \hat{=}\ \bar{\pi},\nonumber \\
\partial_{r}U & = & (\varepsilon+\bar{\varepsilon})-\pi W-\bar{\pi}\bar{W}\ \hat{=}\ 0,\nonumber \\
\partial_{r}X & = & -\pi\xi-\bar{\pi}\bar{\zeta},\\
\partial_{r}\xi & = & -\mu\xi-\bar{\lambda}\bar{\zeta},\nonumber \\
\partial_{r}\zeta & = & -\mu\zeta-\bar{\lambda}\bar{\xi}.\nonumber
\end{eqnarray}
To get the second order derivative of function $U$, we need the first order derivative of connection coefficient $\varepsilon$. This can be achieved by using the second Cartan structure equations, which give
\begin{eqnarray}
-\partial_r(\varepsilon+{\bar\varepsilon})=2|\pi|^2+2Re(\Psi_2)-\frac{R}{12}+2\Phi_{11}.
\end{eqnarray}
Since we consider the Einstein-Maxwell system, we know the scalar curvature vanishes. The above equation also contains the energy-momentum tensor $\Phi_{11}$ of the Maxwell field, so one has to consider the near horizon Maxwell field. Based on the NP formulism of Maxwell equations\cite{SKMHH} and previous analysis, the Maxwell equations on the horizon are
\begin{eqnarray}
\partial_t\Phi_1&\heq& 0,\nonumber\\
\partial_t\Phi_2&\heq&\bar{\delta}\Phi_{1}+2\pi\Phi_{1},\nonumber \\
\nabla_m\Phi_{1}-\partial_{r}\Phi_{0} & \hat{=} & 0,\label{eq:Maxwell}\\
\nabla_m\Phi_{2}-\partial_{r}\Phi_{1} & \hat{=} & 2\mu\Phi_{1}-2\beta\Phi_{2}.\nonumber
\end{eqnarray}

To work out the unknown functions in (\ref{eq:tetrad}), we need the near
horizon behavior of the Ricci tensor component $\Phi_{11}$. It is
given by $\Phi_{11}=2\left|\Phi_{1}\right|^{2}$ where $\Phi_{1}=\frac{1}{2}F_{ab}(l^{a}n^{b}+\bar{m}^{a}m^{b})$
according to \cite{SKMHH}. On the other hand, we know the definitions
of the eletric charge $Q_{e}$ and magnetic charge $Q_{m}$ of a black
hole:
\begin{eqnarray}
Q_{e} & = & \frac{1}{4\pi}\int_{\hat{\Delta}}*F=\frac{1}{2\pi}\int_{\hat{\Delta}}F_{ln}{=\frac{1}{\pi}\int_{{\hat\Delta}}Re({\hat\Phi_1})},\\
Q_{m} & = & \frac{1}{4\pi}\int_{\hat{\Delta}}F=\frac{1}{2\pi}\int_{\hat{\Delta}}F_{\bar{m}m}{=\frac{1}{\pi}\int_{{\hat\Delta}}Im({\hat\Phi_1})},
\end{eqnarray}
in which $\hat{\Delta}$ is a two-dimensional space section surrounding
the black hole. Since $Q_{e}$ and $Q_{m}$ are well defined, $\Phi_{1}$
and $\Phi_{11}$ are regular on the horizon.

Thus we obtain
\begin{eqnarray}
U & = & -[2|{\hat\pi}|^{2}+Re({\hat\Psi}_2)+{\hat\Phi}_{11}]r^{2}+O(r^{3}),\nonumber \\
W & = & {\hat{\bar\pi}}r+O(r^{2}),\label{eq:UWX}\\
X & = & -({\hat\pi}{\hat\xi}+{\hat{\bar\pi}}{\hat{\bar\zeta}})r+O(r^{2}),\nonumber
\end{eqnarray}
{where ``$\ {\hat{\ }}\ $'' means restricting a function on $\Delta$}.
In order to consider the near horizon limit, we need the time dependent
relations of $U,X$ and $W$ further. The commutative relation of
$l$ and $m$ gives $\partial_{t}\xi\heq\partial_{t}\zeta\heq 0.$
{Recall} the IH condition (\ref{eq:1form_poten}),
{ which} means $\partial_{t}\pi\heq 0$, then combining the Bianchi identity and Eq.(\ref{P01}), we get $\partial_{t}\Psi_{2}\heq 0$.
Combining the above results with Eq.(\ref{eq:UWX}), we {know that} the leading order terms of $U,W,X,\xi,\zeta$ in (\ref{eq:tetrad}) are all time independent.

In Einstein-Maxwell case, following the method of ref.\cite{KN_CFT}, we also need to consider the near horizon limit of {the gauge} potential $A$. Because $A$ has gauge freedom, by solving Lorentz gauge condition and the relation between $A$ and $F_{ab}$, one can show that there exist a gauge choice such that the following equation holds,
\begin{eqnarray}
&&A_t\heq 0,\nonumber\\
&&A_r\heq 0,\nonumber\\
&&\tilde{D}^2\tilde{A}\heq \tilde{D}\cdot\tilde{F}-\tilde{d}F_{ln},
\end{eqnarray}
where $\tilde{A}$ is the tangent part of $A$ on the section of horizon, $\tilde{F}$ is the tangent part of $F_{ab}$ on the section of horizon, $\tilde{D}$ is the induced connection on the section of horizon and $\tilde{D}^2$ is the associated Laplace operator. It is clear that the {third} equation is a Laplace equation on $S^2$ so $\tilde{A}$ is fixed by the inner horizon data ${\hat\Phi}_1$ uniquely. From the condition $F_{lm}\heq 0$, one can also find $\partial_t A_m\heq 0$. Since $\Phi_1=F_{ln}+F_{m{\bar m}}$ belongs to inner data of hrizon, one can get the first derivative of $A_t$ on horizon as $\partial_rA_t\heq F_{ln}=Re(\Phi_1)$. From Maxwell equation (\ref{eq:Maxwell}), we also know that $\partial_t\partial_r A_t\heq 0$ is also time independent. So we get the leading term of $A$ are all time independent.

Now the inverse metric could be written as
\begin{eqnarray}
g^{ab} & = & \left(\begin{array}{cccc}
0 & -1 & 0 & 0\\
-1 & f_{1}r^{2}+O(r^{3}) & f_{2}r+O(r^{2}) & \bar{f}_{2}r+O(r^{2})\\
0 & f_{2}r+O(r^{2}) & 2\xi\bar{\zeta} & |\xi|^{2}+|\zeta|^{2}\\
0 & \bar{f}_{2}r+O(r^{2}) & |\xi|^{2}+|\zeta|^{2} & 2\bar{\xi}\zeta
\end{array}\right),
\end{eqnarray}
where
\begin{eqnarray}
f_{1} & = & 6|\pi|^{2}+2Re\Psi_{2}+2\Phi_{11},\nonumber \\
f_{2} & = & 2(\pi\xi+\bar{\pi}\bar{\zeta}).\label{eq:f}
\end{eqnarray}
They are only the functions of the coordinate $\theta$. In NP tetrad,
$l$ and $n$ are future pointing. The outside region of horizon corresponds
to the region $r<0$. We take a transformation $r\rightarrow-r$ for
later convenience. The metric becomes
\begin{eqnarray}
g_{\mu\nu} & = & \left(\begin{array}{ccc}
(h_{ab}f_{2}^{a}f_{2}^{b}-f_{1})r^{2}+O(r^{3}) & -1 & -h_{ab}f_{2}^{a}r+O(r^{2})\\
-1 & 0 & 0\\
-h_{ab}f_{2}^{a}r+O(r^{2}) & 0 & h_{ab}
\end{array}\right)
\end{eqnarray}
with
\begin{eqnarray}
f_{2}^{a} & = & \pi m^{a}+\bar{\pi}\bar{m}^{a},\label{eq:f2a}\\
h_{ab} & = & m_{a}\bar{m}_{b}+m_{b}\bar{m}_{a}.\nonumber
\end{eqnarray}
Here $a,b=\theta,\phi$. $h_{ab}$ is the intrinsic metric of section
$\hat{\Delta}$.

By introducing a coordinate rescaling $r\rightarrow\epsilon\tilde{r},t\rightarrow\frac{\tilde{t}}{\epsilon}$
and taking the near horizon limit $\epsilon\rightarrow0$, we get
the near horizon metric and gauge potential of extreme electrovac
IH.
\begin{eqnarray}
ds^{2} & = & \left(-f_{1}+h_{ab}f_{2}^{a}f_{2}^{b}\right)r^{2}dt^{2}-2dtdr-2h_{ab}f_{2}^{a}rdx^{b}dt+h_{ab}dx^{a}dx^{b}.\\
A & = & -Re({\hat\Phi_1})rdt+\tilde{A}_{\vartheta}d\vartheta+\tilde{A}_{\bar{\vartheta}}d\bar{\vartheta.}\nonumber
\end{eqnarray}
Here the tilde was omitted for conveniences.

{Based on the analysis of this section, an important fact is that the near horizon limit of the metric and {the gauge} potential only depends on the inner horizon data of IH, i.e. $(h,D)$ and ${\hat\Phi}_1$. In other words, any two space-times that each contains an extreme IH will have the same near horizon limit if they share the same inner horizon data.}

\section{The electrovac extreme IH/CFT correspondence}

In last section, the general near horizon metric and gauge field of
extreme electrovac IH have been derived. In this section, we will work
out the explicit form of the metric of axisymmetric electrovac extreme
IHs and compare it with that of the extreme KN spacetime.

It has been proved that all the axi-symmetric electrovac extreme isolated
horizons coincide with the extreme Kerr-Newman event horizon\cite{Uniq_Eextre_IH}.
For axisymmetric extreme IH, there is a vector field $(\partial_{\varphi})^{a}$
that generates the proper symmetry group $O(2)$ of the IH. Instead
of the auxiliary coordinate $\theta$, Lewandowski and Pawlowski introduced
a function $x$. It is defined by $\epsilon_{ab}(\partial_{\varphi})^{b}=2(dx)_{a}$.
The coordinate $x$ is in region $[-\frac{A}{8\pi},\frac{A}{8\pi}]$
where $A$ is equal to the area of $\hat{\Delta}$ and a constant.
In coordinate $(\phi,x)$, the null 2-frame on $\hat{\Delta}$ takes
the following form
\begin{eqnarray}
m & = & \frac{1}{2}(\frac{1}{P}\partial_{x}+iP\partial_{\varphi}).
\end{eqnarray}
 The 1-form $\omega_{a}$ can be decomposed with two functions $U$
and $B$ which are globally defined on $\hat{\Delta}$.
\begin{eqnarray}
\omega & = & \star dU+d\ln B.
\end{eqnarray}
 Here $``\star"$ stands for the intrinsic Hodge dual on section $\hat{\Delta}$.
In NP notations, we have
\begin{equation}
\omega_{a}=-(\varepsilon+\bar{\varepsilon})n_{a}+(\alpha+\bar{\beta})\bar{m}_{a}+(\bar{\alpha}+\beta)m_{a},
\end{equation}
 So the spin coefficient $\pi$ becomes, for extreme IH,
\begin{eqnarray}
\pi & = & -i\bar{\delta}U+\bar{\delta}\ln B.\label{eq:pai}
\end{eqnarray}

In terms of $U,B$ and $P$, {based on the result of Lewandowski and Pawlowski\cite{Uniq_Eextre_IH}}, the solution of axi-symmetric electrovac
extremal IH could be expressed by three real parameters $A,\alpha$
and $\theta_{0}$,
\begin{eqnarray}
P^{2} & = & \frac{4\pi(1+\alpha^{2})}{A}\frac{1+b^{2}x^{2}}{1-(\frac{8\pi}{A}x)^{2}},\nonumber \\
U & = & -\arctan(bx),\label{eq:PUBphi}\\
B & = & (1+b^{2}x^{2})^{1/2},\nonumber \\
\Phi_{1} & = & e^{i\theta_{0}}\sqrt{\frac{\pi}{A}}\frac{2\alpha}{1+\alpha^{2}}\frac{1}{(1\pm ibx)^{2}}.\nonumber
\end{eqnarray}
 Here $b^{2}=\frac{1-\alpha^{2}}{1+\alpha^{2}}(\frac{8\pi}{A})^{2}$,
$\alpha\in[0,1]$, $\theta_{0}\in[0,2\pi)$.

By definition, the electric and magnetic charges are given by the
real and imaginary part of the internal respectively.
\begin{eqnarray}
\frac{1}{4\pi}\int_{\hat{\Delta}}*F+iF & = & e^{i\theta_{0}}\alpha\sqrt{\frac{A}{4\pi}}.
\end{eqnarray}
 Thus $Q_{e}=Q\cos\theta_{0}$ and $Q_{m}=Q\sin\theta_{0}$ where
$\alpha^{2}\frac{A}{4\pi}=Q^{2}$.

Now we could work out the parameters in (\ref{eq:f}). From (\ref{eq:pai})
and (\ref{eq:PUBphi}), we have
\begin{eqnarray}
\pi & = & -i\frac{1}{2P}\frac{b}{1+b^{2}x^{2}}+\frac{1}{2P}\frac{b^{2}x}{1+b^{2}x^{2}},\label{eq:pi2}
\end{eqnarray}
 Combining with (\ref{eq:f2a}), there are
\begin{eqnarray}
h_{ab}f_{2}^{a}dx^{b} & = & \frac{2b^{2}x}{1+b^{2}x^{2}}dx-\frac{1}{P^{2}}\frac{2b}{1+b^{2}x^{2}}d\varphi,\nonumber \\
h_{ab}f_{2}^{a}f_{2}^{b} & = & \frac{2b^{2}}{P^{2}(1+b^{2}x^{2})}.
\end{eqnarray}

We come to $f_{1}$. For electrovac IH, $R=0$. The NP equations and
Einstein-Maxwell equation lead to
\begin{eqnarray}
f_{1} & = & 6\left|\pi\right|^{2}-K+8\left|\Phi_{1}\right|^{2}.\label{eq:f1}
\end{eqnarray}
 Here the Gaussian curvature $K$ of $\hat{\Delta}$ is
\begin{eqnarray}
K & = & \frac{16\pi}{A(1+\alpha^{2})^{2}}\frac{1-3b^{2}x^{2}}{(1+b^{2}x^{2})^{3}}.\label{eq:K}
\end{eqnarray}
 Plugging (\ref{eq:PUBphi},\ref{eq:pi2},\ref{eq:K}) into (\ref{eq:f1}),
we finally get
\begin{eqnarray}
f_{1} & = & \frac{3Ab^{2}}{8\pi(1+\alpha^{2})}\frac{1-(\frac{8\pi}{A}x)^{2}}{(1+b^{2}x^{2})^{2}}-\frac{16\pi}{A(1+\alpha^{2})^{2}}\frac{1-3b^{2}x^{2}}{(1+b^{2}x^{2})^{3}}\nonumber \\
 &  & +\frac{8\pi}{A}(\frac{2\alpha}{1+\alpha^{2}})^{2}\frac{1}{(1+b^{2}x^{2})^{2}}.
\end{eqnarray}
 The metric could be reorganized as
\begin{eqnarray}
ds^{2} & = & -f_{1}r^{2}dt^{2}-2dtdr\label{eq:metric_IH}\\
 &  & +2P^{2}(dx-\frac{1}{P^{2}}\frac{b^{2}x}{1+b^{2}x^{2}}rdt)^{2}\nonumber \\
 &  & +\frac{2}{P^{2}}(d\varphi+\frac{b}{1+b^{2}x^{2}}rdt)^{2}.\nonumber
\end{eqnarray}

Let us turn to the electromagnetic field. From (\ref{eq:PUBphi}),
we get the gauge field strength on $\Delta$.
\begin{eqnarray}
F_{rt} & = & \frac{4\pi}{A(1+\alpha^{2})}\frac{Q_{e}(1-b^{2}x^{2})+2Q_{m}bx}{(1+b^{2}x^{2})^{2}},\\
F_{x\varphi} & = & \frac{8\pi}{A(1+\alpha^{2})}\frac{Q_{m}(1-b^{2}x^{2})-2Q_{e}bx}{(1+b^{2}x^{2})^{2}}.\nonumber
\end{eqnarray}
Since the gauge field is time-independent and
rotationally symmetric, the components of gauge potential $A$ are only
functions of $x,r$. On the other hand, gauge condition $A_{r}=0$
could always be chosen. $A_{x}$ is only the functions of $x$ on
$\Delta$ which could also be gauge fixed to $A_{x}=0$. So we get
\[
A_{r}=0,\ \ \ \ \ A_{x}=0.
\]
\begin{eqnarray}
A_{t} & = & \frac{4\pi}{A(1+\alpha^{2})}\frac{Q_{e}(1-b^{2}x^{2})-2Q_{m}bx}{(1+b^{2}x^{2})^{2}}r+O(r^{2}),\label{eq:A_EIH}\\
A_{\varphi} & = & \frac{4\pi}{Ab(1+\alpha^{2})}\frac{Q_{e}(1-b^{2}x^{2})+2Q_{m}bx}{1+b^{2}x^{2}}+O(r).\nonumber
\end{eqnarray}

By introducing a coordinate rescaling $r\rightarrow\epsilon\tilde{r},t\rightarrow\frac{\tilde{t}}{\epsilon}$
and taking the near horizon limit $\epsilon\rightarrow0$, we get
the near horizon geometry and gauge field of extreme electrovac IH.
They have the same form as (\ref{eq:metric_IH},\ref{eq:A_EIH}) if
the tilde was omitted.

Now the metric of near horizon extreme IH is determined by two parameters:
the horizon area $A$ and charge $Q$. The metric of extreme KN is
also determined by two parameters: the mass $M$ and angular momentum
$a$. For extreme KN, $M^{2}=a^{2}+Q^{2}$ and the area of event horizon
$A=4\pi(M^{2}+a^{2})$. Thus the extreme KN could also be described
by the horizon area $A$ and the electric charge $Q$. Based on the
uniqueness theorem, the near horizon geometry of extreme IH and extreme
KN should be the same. In fact, if the coordinate $x\equiv\frac{A}{8\pi}\cos\theta$
was introduced in extreme KN, and the two parameters $M$ and $a$
were replaced by $A=4\pi(M^{2}+a^{2})$ and $\alpha^{2}=\frac{4\pi}{A}Q^{2}=\frac{M^{2}-a^{2}}{M^{2+a^{2}}}$,
we will find that the near horizon limit metric of extreme KN is exactly
the same as that of extreme IH. To make this clearer, we proceed to
make a coordinate transformation for line element (\ref{eq:NHEKN}).
\begin{eqnarray}
\rho & = & \frac{M^{2}+a^{2}\cos^{2}\theta}{M^{2}+a^{2}}r,\nonumber \\
v & = & (M^{2}+a^{2})(t+\frac{1}{r}),\label{eq:Poin2Bondi}\\
\varphi & = & \phi+\frac{2Ma}{M^{2}+a^{2}}\ln r.\nonumber
\end{eqnarray}
 Then the metric of near horizon limit extreme KN becomes \begin{CJK}{GB}{}{
\begin{eqnarray}
ds^{2} & = & -\frac{(M^{2}+a^{2}\cos^{2}\theta)^{2}+4a^{4}\cos^{2}\theta\sin^{2}\theta}{(M^{2}+a^{2}\cos^{2}\theta)^{3}}\rho^{2}dv^{2}-2dvd\rho\label{eq:Bondi_KN}\\
 &  & +(M^{2}+a^{2}\cos^{2}\theta)(d\theta+\frac{2a^{2}\cos\theta\sin\theta}{(M^{2}+a^{2}\cos^{2}\theta)^{2}}\rho dv)^{2}\nonumber \\
 &  & +\frac{(M^{2}+a^{2})^{2}\sin^{2}\theta}{M^{2}+a^{2}\cos^{2}\theta}(d\varphi+\frac{2Ma}{(M^{2}+a^{2})(M^{2}+a^{2}\cos^{2}\theta)}\rho dv)^{2}.\nonumber
\end{eqnarray}
}\end{CJK}\inputencoding{latin9} This is the Bondi coordinate for
KN. It is easy to shown that
\begin{eqnarray}
\frac{(M^{2}+a^{2}\cos^{2}\theta)^{2}+4a^{4}\cos^{2}\theta\sin^{2}\theta}{(M^{2}+a^{2}\cos^{2}\theta)^{3}} & = & f_{1},\\
\frac{(M^{2}+a^{2})^{2}\sin^{2}\theta}{M^{2}+a^{2}\cos^{2}\theta} & = & \frac{2}{P^{2}}.\nonumber
\end{eqnarray}
Thus the near horizon metric of extreme (\ref{eq:Bondi_KN}) is the
same as that of extreme IH (\ref{eq:metric_IH}).

Besides, the gauge field (\ref{eq:gauge_field}) under coordinate
transformation (\ref{eq:Poin2Bondi}) becomes
\begin{eqnarray}
A & = & f(\theta)(d\varphi+\frac{k\rho}{M^{2}+a^{2}\cos^{2}\theta}dv),
\end{eqnarray}
which is exactly the same as (\ref{eq:A_EIH}).

So it has been proved that the near horizon limit of a general electrovac axisymmetric
extreme IH is exactly the same as {that of} the extreme KN black hole.
According to the construction of the Bondi-like coordinate,
$\partial_{t}$ is an asymptotic Killing vector of IH up to some terms
of high order. This enables us to take Fourier transformation to define
the Frolov temperature, which is similar to the case of Kerr spacetime.
It has been shown that IHs have the thermodynamic properties similar
to traditional black hole horizons. Hawking radiation is emitted by
IHs as well. So IHs also have entropy. On the other hand, the correspondence
between KN-AdS and CFT was uncovered in \cite{KN_CFT}. Since we have
shown that the near horizon metric and gauge field of electrovac extreme
IHs are the same as those of extreme KN, the electrovac extreme IH/CFT
reduces to the case of KN-AdS/CFT by vanishing the cosmological constant,
as we have presented in section 2. Thus the electrovac extreme IHs
are holographically dual to a chiral two-dimensional CFT with central
charge $c=12J$. The macroscopic Bekenstein-Hawking entropy of electrovac
extreme IHs can be reproduced by computing the microscopic entropy
from the Cardy formula for the two-dimensional conformal field theory.

\section{Summary and discussions}

The Kerr/CFT correspondence is generalized to the electrovac axisymmetric
extreme IH/CFT correspondence. The macroscopic Bekenstein-Hawking
entropy of extreme IH is reproduced by computing the microscopic entropy
for the dual two-dimensional conformal field theory.

We first reviewed the near horizon geometry of KN spacetime, then
derived the general near horizon limit of {spacetimes containing} IHs.
It was shown that, in the near horizon limit, any two spacetimes each contains an extreme IH
will have the same near horizon structure if they have the same inner horizon data. For axisymmetic
electrovac extreme IH, the uniqueness theorem states that it coincides with the extreme
Kerr-Newman event horizon under the near horizon limit.
We built the near horizon limit of {spacetimes containing}
IHs explicitly and found that it is exactly the same as
the near horizon limit of KN spacetime after a coordinate transformation
between the Bondi-like coordinates and Poincare{-type} coordinates{,
where the explicit form of this coordinate transformation is found}.
Then the Kerr/CFT correspondence was generalized to the
electrovac extreme IH/CFT correspondence.

As stressed in Section 3, the IH geometry is intrinsically determined.
This implies that the microscopic origin of IH entropy is determined
by the intrinsic geometry. {We expect that further study on this
aspect will more clearly outline a geometric picture of the Kerr/CFT correspondence.}

{To be even more general, the electrovac spacetimes containing IHs with cosmological constant can be considered,
whose near horizon geometry is expected to be exactly the same as that of the KN-(A)dS spacetime. In order to do that,
however, one needs to generalize Lewandowski and Pawlowski's uniqueness theorem to the case with cosmological constant,
which will be left for future works.}

\section*{Acknowledgments}
This work is partly supported by the Natural Science Foundation of China
under Grant Nos. 11175245, 11075206.


\begin{thebibliography}{10}
\bibitem{Entropy_5D}A. Strominger and C. Vafa, Phys. Lett. B 379
(1996) 99. {[}hep-th/9601029{]}.

\bibitem{Entropy_4D}J.M. Maldacena and A. Strominger, Phys. Rev.
Lett. 77 (1996) 428. {[}hep-th/9603060{]}.

\bibitem{AdS3_CFT}J.D. Brown and M. Henneaux, Commun. Math. Phys.
104, 207 (1986).

\bibitem{CFT_1}A. Strominger, JHEP 9802 (1998) 009. {[}hep-th/9712251{]}.

\bibitem{Kerr_CFT}M. Guica, T. Hartman, W. Song, and A. Strominger,
Phys. Rev. D. 80. 124008. {[}arXiv:0809.4266{]}.

\bibitem{AOT_1KK}T. Azeyanagi, N. Ogawa and S. Terashima, JHEP 0904, 061, 2009.
{[}arXiv:0811.4177{]}.

\bibitem{BMPV}H. Isono, T.-S. Tai and W.-Y. Wen, Int. J. Mod. Phys. A 24, 5659-5668, 2009.
arXiv:0812.4440 {[}hep-th{]}.

\bibitem{Godel}J.-J. Peng and S.-Q. Wu, Phys. Lett. B 673, 216-219, 2009.
arXiv:0901.0311 {[}hep-th{]}

\bibitem{Gauge_SRG}D.D.K. Chow, M. Cvetic, H. Lu and C.N. Pope, Phys. Rev. D 79, 084018, 2009.

\bibitem{Heterotic}A.M. Ghezelbash, JHEP 0908, 045, 2009. arXiv:0901.1670
{[}hep-th{]}.

\bibitem{Central_Charges}G. Comp\`ere, K. Murata and T. Nishioka, JHEP
0905, 077, 2009. arXiv:0902.1001 {[}hep-th{]}.

\bibitem{divers_dim}H. Lu, J. Mei and C.N. Pope, JHEP 0904, 054, 2009.
arXiv:0811.2225 {[}hep-th{]}.

\bibitem{LMPV_SRG}H. Lu, J. Mei, C.N. Pope and J. Vazquez-Poritz,
Phys. Lett. B 673 (2009) 77-82. arXiv:0901.1677 {[}hep-th{]}

\bibitem{Wald_Entropy}C. Krishnan and S. Kuperstein, Phys. Lett. B 677, 326-331, 2009.
arXiv:0903.2169 {[}hep-th{]}.

\bibitem{Min_SRG}C.-M. Chen and J.E. Wang, Class. Quant. Grav. 27 (2010) 075004.
 arXiv:0901.0538 {[}hep-th{]}.

\bibitem{RN_CFT}M.R. Garousi and A. Ghodsi, Phys. Lett. B687 (2010) 79-83. arXiv:0902.4387 {[}hep-th{]}.

\bibitem{Compere}G. Comp\`ere, Living Rev. Relativity {\bf 15} (2012), 11. [Online Article]:
http://www.livingreviews.org/lrr-2012-11

\bibitem{BKLS}I. Bredberg, C. Keeler, V. Lysov and A. Strominger, Cargese Lectures on the Kerr/CFT Correspondence,
Nucl. Phys. Proc. Suppl. 216 (2011) 194.

\bibitem{KN_CFT}T. Hartman, K. Murata, T. Nishioka, and A. Strominger,
arXiv:0811.4393 {[}hep-th{]}.

\bibitem{IH_CFT}X.-N. Wu, Y. Tian, Phys. Rev. D 80, 024014(2009).
arXiv:0904.1554 {[}hep-th{]}.

\bibitem{Ashtekar_1}A. Ashtekar, C. Beetle and S. Fairhurst, Class. Quant.
Grav. 16, L1-L7, 1999. arXiv:gr-qc/9812065.

\bibitem{IH_DH}A. Ashtekar and B. Krishnan, Living Rev. Relativity
{\bf 7} (2004) 10. [Online Article]:
http://www.livingreviews.org/lrr-2004-10. arXiv: gr-qc/0407042.

\bibitem{Lewandowski}J. Lewandowski, Class. Quant. Grav. 17 (2000)
L53-L59. arXiv:gr-qc/9907058.

\bibitem{LP02}J. Lewandowski and T. Pawlowski, Int. J. Mod. Phys. D11 (2002) 739-746.

\bibitem{Uniq_Eextre_IH}J. Lewandowski, T. Pawlowski, Class. Quant. Grav.
20 (2003) 587-606. arXiv:gr-qc/0208032.

\bibitem{Horowitz}J.M. Bardeen and G.T. Horowitz, Phys. Rev. D 60
(1999) 104030. arXiv:hep-th/9905099.

\bibitem{KL13}H. K. Kunduri and J. Lucietti, Living Rev. Relativity, {\bf 16}, (2013), 8. [Online Article]: http://www.livingreviews.org/lrr-2013-8

\bibitem{Bondi-coor}H. Friedrich, Proc. Roy. Soc. Lond. A 378 (1981)
169-184, 401-421.

\bibitem{GeoIH}A. Ashtekar, C. Beetle, J. Lewandowski, Class. Quant.
Grav. 19, 1195-1225, 2002. arXiv:gr-qc/0111067.

\bibitem{SKMHH} H. Stephani, D. Kramer, M. A. H. MacCallum, C. Hoenselaers,
E. Herlt, {\it Exact Solutions of Einstein's Field Equations}, Cambridge
University Press 2003.

\end{thebibliography}
\end{document}